\documentclass[sigconf]{acmart}
\usepackage{booktabs}
\usepackage{tabularx}
\usepackage{array}
\usepackage{multirow}
\usepackage{xcolor}
\usepackage{longtable}
\usepackage{array}
\usepackage{float}

 \usepackage{float}

\usepackage{colortbl}
\usepackage{enumitem}
\usepackage{fontawesome5}
\usepackage{subfig} 
\usepackage{makecell}  
\usepackage{booktabs}  
\usepackage{tikz}
\usetikzlibrary{arrows.meta, positioning}
\AtBeginDocument{%
  }

\copyrightyear{2026}
\acmYear{2026}
\setcopyright{cc}
\setcctype{by}
\acmConference[IH '26]{Interactive Health Conference}{July 05--08, 2026}{Porto, Portugal}
\acmBooktitle{Interactive Health Conference (IH '26), July 05--08, 2026, Porto, Portugal}
\acmDOI{10.1145/3786579.3804910}
\acmISBN{979-8-4007-2422-0/2026/07}
\begin{document}
\title{Designing for Patient Voice in Interactive Health}


\author{Yuhao Sun}
\email{yuhao.sun@lancaster.ac.uk}
\orcid{0000-0002-4053-6032}
\affiliation{%
  \institution{Lancaster University}
  \city{Lancaster}
  \state{England}
  \country{United Kingdom}
}
\affiliation{%
  \institution{University of Edinburgh}
  \city{Edinburgh}
  \state{Scotland}
  \country{United Kingdom}
}


\begin{abstract}

Interactive Health (IH) research increasingly engages patients through participatory and user-centred approaches. However, patients' lived experiences are typically treated more as data to be analysed than as knowledge in their own right. In this paper, I argue that `patient voice' in the field of IH is both an inclusion issue and an epistemic one. More specifically, it concerns how experiential accounts are recognised and circulated. I examine how methodological conventions, authorship norms, review criteria, and publication formats tend to position patients as participants rather than as authors of evidence. Looking to patient-partnered practices in medical publishing, including \textit{The BMJ}, \textit{JAMA}, and \textit{British Journal of Sports Medicine}, I outline a possible infrastructural pathway for supporting patient-authored or patient-led experiential contributions within the field. I present this as a design probe to surface assumptions and trade-offs. I end this paper by inviting the IH community to reflect on how its knowledge infrastructures might accommodate experiential evidence alongside established research forms.

\end{abstract}





\maketitle

\section{Introduction}

Interactive Health (IH)\footnote{In this paper, the term ``Interactive Health (IH)'' is adopted as a field-level descriptor rather than referring to the conference venue. I use IH to denote the broader body of research, practices, and scholarly communities working at the intersection of Human-Computer Interaction (HCI) and health, including work that predates the formalisation of dedicated IH venues.} has rapidly matured as a research space at the intersection of health, care, and interactive technologies \cite{avellino2025envisioning,blandford2018seven,waddell2024leveraging,epstein2023symposium}. Across topics such as AI-enabled clinical decision support, self-tracking systems, digital therapeutics, and health data infrastructures, the community has produced substantial advances in how health-related practices are mediated, augmented, and reorganised through interaction, strengthening dialogue between HCI, medicine, and digital health research.

However, as the field of IH develops its technical and methodological set, an underlying asymmetry remains in whose knowledge is structurally centred. IH research is commonly conducted through researcher-led studies, in which patients are involved as participants, data sources, or end users \cite{blandford2018seven,waddell2024leveraging,epstein2023symposium}. Within these arrangements, lived experience is \textit{typically} translated into themes, metrics, or design requirements through academic analytic frameworks. What is less visible here is a limited epistemic positioning: while patients may contribute substantively, this contribution might be mediated and not reflected in authorship.

In this paper, I argue that `patient voice' in the field of IH should be understood not as a matter of inclusion or representation alone, but as a question of epistemic completeness -- that is, whether experiential knowledge about interactive health systems is recognised and shared as knowledge in its own right. This reframes patient voice as experiential evidence and shifts attention from engaging patients in research towards designing the epistemic infrastructures through which such knowledge can be shared and valued. To develop this argument, I examine how patient voice is currently positioned within IH research and identify structural features that tend to treat experiential knowledge as data rather than as authored knowledge. Drawing on practices from participatory health research and patient-partnered medical publishing, I outline an infrastructural framing through which patient-authored or patient-led experiential accounts might enter IH's scholarly discourse, as a design probe.

\section{Patient Voice as an Epistemic Category}

First, what is the `patient voice'? The notion of patient voice appears across health research, policy discourse, and design practice, often associated with public engagement, patient and public involvement, participatory design, or user-centred approaches \cite{gilmore2019value,perfetto2017value,johnson2016patient,leblanc2017patient}. Within these framings, `patient voice', or more intuitively, patient perspectives, are \textit{typically} valued as a way to improve relevance, acceptability, and usability of interventions and systems. For example, patients contribute feedback, share their needs, or help shape design directions. Vines et al. \cite{vines2013configuring} point out that participation is always configured through the structures and decisions of design and research practice. In many health research contexts, patients are involved in processes defined and interpreted by academic or clinical actors. In this sense, `voice' functions as an input to knowledge production but not as knowledge itself.

Meanwhile, Bardzell \cite{bardzell2010feminist} reminds us that knowledge is situated and that marginalised or embodied perspectives are not anecdotal residues of research but sites of knowledge production in their own right. Following this perspective, experiential accounts in IH can be understood as forms of situated knowledge produced through lived engagement with health systems and technologies. Patients develop situated knowledge about symptoms, treatment trade-offs, system navigation, and the everyday consequences of clinical and technological decisions. Is this knowledge just anecdotal? Rather, it is produced through sustained engagement with illness, care infrastructures, and health technologies over time. Suchman \cite{suchman2007human} also argues that knowledge and action are always situated within ongoing relations between people and systems, so patient voice can be understood as experiential evidence: a way of knowing grounded in embodied, relational, and temporal experience of health systems.

I argue this shift from voice-as-input to voice-as-evidence carries epistemic implications for the field of IH. If experiential knowledge is recognised as evidence, then the question is no longer only how to involve patients in studies, but how to position patient-authored or patient-led accounts within the field's knowledge structures. This reframing shifts the discussion from participation to epistemic status: whose accounts count, under what criteria, and through what formats \cite{rogers2012hci,harrison2007three}. In doing so, `patient voice' becomes less a stakeholder label and more an epistemic category that challenges how IH defines valid knowledge about interactive health systems \cite{gaver2012should,stolterman2010concept}. In the following subsections, I explore this epistemic reframing by clarifying who ``patients'' include in IH, examining structural barriers in current knowledge practices, and looking at how other health venues have made space for experiential contributions.

\subsection{Defining Patient Voice Broadly}

While the term ``patient''\footnote{The term ``patient'' itself is debated. Erlend Hem's article ``Patient, client, user or customer?'' discusses how alternative labels carry different historical, legal, and relational implications \cite{hem2013patient}. Hem argues that although the term ``patient'' has its flaws, substitutes such as ``client,'' ``user,'' or ``customer'' introduce their own conceptual and normative problems, particularly in how they frame responsibility, power, and the doctor-person relationship.} often evokes individuals receiving clinical care \cite{neuberger1999we,eisenberg1980makes}, in IH contexts it refers more broadly to people whose lives are shaped through sustained engagement with health systems, conditions, and technologies \cite{mamykina2010constructing,sun2025human,bhat2023we,vines2013making,sun2025future,sun2026how}. This includes individuals living with chronic or episodic conditions \cite{mamykina2010constructing}, those navigating preventive care or risk management in contexts such as precision medicine \cite{sun2025human,sun2025future,sun2026how}, caregivers and family members involved in care work \cite{bhat2023we}, and people who interact with health infrastructures outside formal clinical encounters \cite{vines2013making}. What unites these groups is a relational position: they experience health systems and technologies as those who must live with their consequences over time.

Recognising this broader scope prevents `patient voice' from being reduced to a clinical or disease-specific category. Instead, it foregrounds a common experiential condition: living within health-related socio-technical systems whose decisions, classifications, and data practices shape everyday life. In IH research, for example, these experiences may involve algorithmic risk scores, self-tracking feedback loops, remote monitoring infrastructures, or digital care pathways. By defining ``patient'' in this relational sense, `patient voice' becomes relevant across the breadth of IH topics rather than confined to particular medical domains.

\subsection{Barriers in Current Structures}

Despite growing recognition of patient engagement in IH research, the ways knowledge is structured, reviewed, and circulated in the field of IH can make it difficult for experiential knowledge to enter the field as knowledge in its own right. As Blandford et al. \cite{blandford2018seven} observe, IH research brings together distinct methodological lifecycles and publication cultures from HCI and health research, each carrying particular assumptions about evidence and contribution. I argue that within such configurations, barriers are embedded in academic norms that shape what counts as a legitimate contribution. In the following, I examine these barriers across four dimensions.

First, \textbf{methodological framing.} IH research has tended to prioritise contributions framed through established research designs, analytic procedures, and generalisable findings \cite{bowman2023using}. Experiential accounts, by contrast, are often expected to be translated into themes, models, or design implications before they become publishable. While this translation supports comparability and abstraction, it can also detach lived experience from the contexts, ambiguities, and temporalities through which it gains meaning. As a result, experiential knowledge may be present in IH studies, but primarily as material to be analysed, rather than as authored accounts with epistemic standing \cite{bardzell2010feminist}.

Second, \textbf{authorship and voice.} Academic authorship norms further shape whose voice appears as knowledge. Publications are conventionally authored by researchers, with patients acknowledged as participants, advisors, or collaborators. In participatory co-design research, the final narrative is \textit{typically} structured through academic discourse and interpretive authority \cite{vines2013configuring,bardzell2010feminist,le2015strangers}. This arrangement does not negate patient involvement, but it positions experiential knowledge within researcher-mediated representation \cite{simonsen2013routledge}. Therefore, opportunities for patient-authored or patient-led accounts remain rare as existing formats are not designed to accommodate them \cite{le2015strangers}. This does not assume that all patients seek or desire authorship, but rather acknowledges those who wish to contribute experiential knowledge as authored accounts.

Third, \textbf{review criteria and epistemic expectations.} Broadly, peer review criteria have historically reflected expectations regarding contribution, evidence, and methodological grounding \cite{wobbrock2016research}. At major venues where IH research is published (e.g., CHI, CSCW, or DIS), submissions are commonly evaluated in relation to clarity of research design, analytic coherence, and the articulation of broader implications for the field. These evaluative norms have largely evolved within researcher-led traditions of inquiry and academic training. While such criteria provide shared standards for assessing scholarship, experiential narratives that do not readily align with established forms of methodological framing may be perceived as anecdotal or insufficiently analytic. These judgments reflect how established evaluative criteria may not readily accommodate forms of knowledge grounded in lived experience. However, when such criteria are applied uniformly, they can inadvertently marginalise forms of knowledge grounded in lived experience.

Finally, \textbf{format and genre constraints.} Publication formats and genre conventions shape how contributions are expected to appear. For many empirical contributions, abstracts, method sections, and results organised around analytic categories remain the dominant mode of scholarly presentation, often reflecting conventions derived from formats such as IMRAD \cite{sollaci2004introduction}. Prior work at CHI also highlights that community-specific writing and citation norms play a role in determining what forms of scholarly presentation are rewarded or visible \cite{pohl2019we}. In addition, submission infrastructures, such as prescribed templates, formatting requirements (e.g., the ACM Master Article Submission templates), and conference management systems (e.g., the PCS Submission System), are designed around assumptions of academic authorship and familiarity with scholarly publishing practices. These combined expectations offer limited flexibility for narrative, reflective, or longitudinal experiential accounts. Without dedicated formats, guidance, or alternative submission routes, such contributions may struggle not only in how they are written, but also in navigating publication procedures, rendering them less legible within the field's established forms of scholarship.

Overall, these structures do not operate through the intentional exclusion of patient voice. Rather, they create conditions in which experiential knowledge must be reshaped to align with researcher-centric epistemic norms before it can circulate as recognised knowledge. The issue, therefore, is not a lack of goodwill or engagement, but a structural positioning in which experiential knowledge must first be translated into researcher-legible form before it can circulate as recognised knowledge.

\subsection{Learning from Other Health Venues}

To understand how experiential knowledge can be institutionally supported within scholarly publishing, I turn to examples from other areas of health that have structured space for first-person accounts. Below, I briefly introduce relevant sections in \textit{The BMJ}, \textit{JAMA}, and the \textit{British Journal of Sports Medicine}. These examples demonstrate that experiential accounts can be integrated into formal knowledge infrastructures without necessarily conforming to conventional research article formats.

In general medical publishing, first, \textit{The BMJ} provides a notable example through its patient-led series \textit{What Your Patient Is Thinking}\footnote{\hyperlink{https://www.bmj.com/about-bmj/resources-authors/article-types/WYPIT}{https://www.bmj.com/about-bmj/resources-authors/article-types/WYPIT}} \cite{snow2015you}. These short narrative pieces, typically around 650 words, are written primarily by patients and carers. They are explicitly framed around what clinicians might learn from lived experience. Rather than aiming for generalisable findings, the series positions experiential accounts as a source of practical insight into communication, care processes, and the consequences of clinical decisions. Editorially, these contributions are developmentally shaped by patient and clinical editors, emphasise plain language, limit academic conventions such as extensive referencing, and are integrated into the journal's formal publication structure. Submissions are handled through direct contact with the series editor, and prospective authors may submit an outline or pitch before developing a full piece.

In addition, \textit{JAMA (The Journal of the American Medical Association)} has long included the section \textit{A Piece of My Mind}\footnote{\hyperlink{https://jamanetwork.com/journals/jama/pages/instructions-for-authors\#SecAPieceofMyMind}{https://jamanetwork.com/journals/jama/pages/instructions-for-authors\#SecAPieceofMyMind}} since the first essay in 1984 \cite{malani2020forty}, which publishes short personal essays reflecting on experiences in medicine, including patient-physician relationships and the human dimensions of care. These pieces, limited to 1,600 words, centre on first-person or professionally situated narratives, yet are fully embedded within the journal's formal publication system. Submissions follow standard manuscript procedures, and must comply with strict ethical requirements, including documented patient consent when individuals are identifiable. Fictionalised or composite accounts are not permitted.

In more specialised areas of medicine, such as sport and exercise medicine, the \textit{British Journal of Sports Medicine (BJSM)} provides a further example through its \textit{Patient Voices}\footnote{\hyperlink{https://bjsm.bmj.com/pages/authors\#patient\_voices}{https://bjsm.bmj.com/pages/authors\#patient\_voices}} section \cite{ahmed2018creating}. It focuses on first-person accounts of health and performance experiences in sport, exercise, and physical activity, explicitly aiming to reflect a global and lifespan-diverse community. The section actively encourages submissions from individuals and communities underrepresented in sport and exercise medicine literature, including people with disabilities, those with chronic conditions, LGBTQIA2S+ communities, and contributors from low- and middle-income settings. Editorially, the genre is specified as short pieces (around 1,000 words) without abstracts, with limited references and optional visual elements, submitted through a dedicated editorial contact. Guiding questions provided to authors frame lived experience around themes such as injury, mental health, access to care, social determinants, and community belonging. Submissions undergo editorial review and may first appear on the \textit{BJSM} blog, with selected pieces progressing to journal publication.

Across these venues, first-person contributions engage with questions of measurement, mediation, identity, and infrastructural care. Some of these concerns resonate closely with core themes in the field of IH. In \textit{The BMJ}, for example, one piece describes living with white coat syndrome, focusing on the challenges of explaining fluctuating blood pressure readings and the role of medical records in stabilising health data interpretation across clinical encounters \cite{Lawriem443How}. In \textit{JAMA}, essays have explored smartphone-mediated communication at the end of life \cite{jenq2025til}, experiences of exclusion within online health communities shaped by professional identity \cite{wexler2024my}, and the deployment of emergency telehealth during natural disasters \cite{heyworth2018sharing}. In \textit{BJSM}, a narrative account of rehabilitation from the perspective of a hijabi athlete examines recovery as a process entangled with cultural, religious, and identity-based negotiations rather than solely physiological restoration \cite{amer2024rehabilitation}.

\section{An Open Structural Proposal}
\label{sec3-proposal}

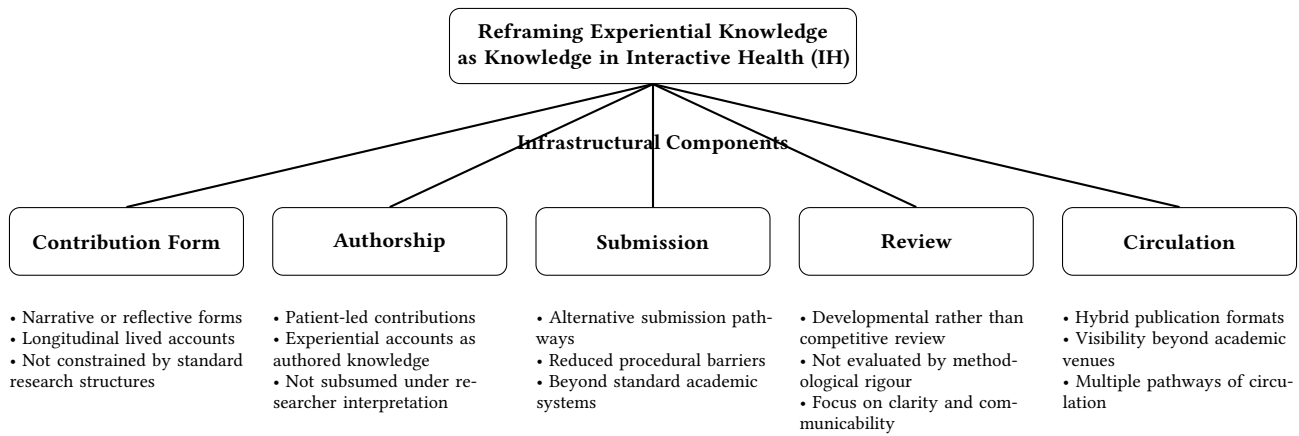
\begin{figure*}[t]
\centering
\begin{tikzpicture}[
    x=1cm, y=1cm,
    box/.style={
        draw,
        rounded corners,
        align=center,
        font=\small,
        minimum height=0.9cm,
        text width=2.9cm
    },
    topbox/.style={
        draw,
        rounded corners,
        align=center,
        font=\small\bfseries,
        minimum height=1cm,
        text width=5.2cm
    },
    labelstyle/.style={
        align=center,
        font=\small\bfseries
    },
    textblock/.style={
        align=left,
        font=\footnotesize,
        text width=3.1cm
    },
    line/.style={thick}
]

\node[topbox] (top) at (0,0) {Reframing Experiential Knowledge\\as Knowledge in Interactive Health (IH)};

\node[labelstyle] at (0,-1.3) {Infrastructural Components};

\node[box] (c1) at (-7,-2.6) {\textbf{Contribution Form}};
\node[box] (c2) at (-3.5,-2.6) {\textbf{Authorship}};
\node[box] (c3) at (0,-2.6) {\textbf{Submission}};
\node[box] (c4) at (3.5,-2.6) {\textbf{Review}};
\node[box] (c5) at (7,-2.6) {\textbf{Circulation}};

\draw[line] (top.south) -- (c1.north);
\draw[line] (top.south) -- (c2.north);
\draw[line] (top.south) -- (c3.north);
\draw[line] (top.south) -- (c4.north);
\draw[line] (top.south) -- (c5.north);

\node[textblock, below=0.35cm of c1] {%
• Narrative or reflective forms\\
• Longitudinal lived accounts\\
• Not constrained by standard research structures
};

\node[textblock, below=0.35cm of c2] {%
• Patient-led contributions\\
• Experiential accounts as authored knowledge\\
• Not subsumed under researcher interpretation
};

\node[textblock, below=0.35cm of c3] {%
• Alternative submission pathways\\
• Reduced procedural barriers\\
• Beyond standard academic systems
};

\node[textblock, below=0.35cm of c4] {%
• Developmental rather than competitive review\\
• Not evaluated by methodological rigour\\
• Focus on clarity and communicability
};

\node[textblock, below=0.35cm of c5] {%
• Hybrid publication formats\\
• Visibility beyond academic venues\\
• Multiple pathways of circulation
};

\end{tikzpicture}

\caption{A conceptual framing of an alternative knowledge infrastructure in IH, outlining key infrastructural components through which such accounts can be expressed, reviewed, and shared as epistemically valid contributions. The figure summarises Section \ref{sec3-proposal}.}
\label{fig:ih-framework}
\end{figure*}

To make the discussion about patient voice in the field of IH more concrete, I present a draft framing that resembles what a call for contributions or author guidance for experiential accounts might look like. Figure \ref{fig:ih-framework} synthesises these dimensions discussed in this section. As previously introduced, I position this as a design probe more than a track proposal: an outline of how such a space could be structured in order to surface the assumptions, values, and trade-offs involved. By outlining a possible format and process, I aim to highlight the infrastructural adjustments that could support the inclusion of experiential knowledge within IH's scholarly discourse.

One possible framing would invite submissions that describe lived experiences of interacting with health-related technologies, infrastructures, or data systems over time. These might include experiences of remote monitoring, risk prediction tools, self-tracking systems, digital care pathways, or algorithmic decision-support systems. The focus here shifts from systems as designed and evaluated to systems as lived with, acknowledging that many consequences of interactive health technologies emerge in longitudinal and everyday contexts that are not fully visible in controlled studies. In doing so, these contributions would be positioned as accounts of how socio-technical arrangements are encountered, interpreted, and negotiated in practice. 

In this framing, contributions could take narrative, reflective, or longitudinal forms without being required to follow conventional research structures such as predefined methods or experimental protocols. Here, narrative functions as a mode of situated analysis, conveying temporal progression, ambiguity, affect, and relational dynamics integral to understanding system consequences. Brief guiding prompts may support authors in framing their accounts. To maintain accessibility while preserving coherence, submissions might be significantly shorter than standard papers, with a limited number of references and optional display items such as timelines, diary excerpts, or artefacts documenting everyday interaction. These materials would function as forms of evidence, making visible dimensions of interactional life that are not easily captured through conventional analytic frameworks. A brief standfirst could accompany the piece in lieu of a conventional abstract.

Authorship can be framed to accommodate both single-authored patient contributions and co-authored work between patients and other collaborators. This does not assume that all patients wish to be named as authors. Instead, it opens space for those who wish to contribute experiential knowledge as authored accounts, while also allowing for forms of anonymous or pseudonymous contribution where preferred. In co-authored submissions, collaborators may provide contextualisation or theoretical linkage to IH discourse. However, the experiential account would remain central rather than being subsumed within collaborator-led interpretation. In this way, authorship functions as an epistemic signal, indicating that lived experience is not just data for analysis but a site of knowledge production. Language expectations would similarly require careful consideration: while English may remain the primary publication language, mechanisms for translation support or editorial assistance could help ensure that contributors who are less confident in academic English are not excluded. 

Operationally, submission and review processes would need to balance accessibility with scholarly accountability. For instance, submissions might be accepted through a dedicated email channel rather than the full conference management pipeline, thus lowering procedural barriers. Review could be handled by a small number of associate chairs or an appointed committee familiar with participatory and patient-partnered research, potentially operating on a rolling basis rather than tied to a single annual deadline. Lived-experience writing does not always align with conference cycles, and review could therefore be framed less as competitive ranking and more as developmental assessment. Outcomes might range from acceptance to revision or rejection, with feedback focused on clarity, relevance to the field of IH, and communicability. Over time, stewardship of this space might require oversight from a standing group or steering function.

Beyond process considerations, the positioning of these contributions within the broader IH ecosystem would be equally important. Accepted pieces might first be published on an IH-affiliated blog or online platform dedicated to these accounts, or curated in a zine format, creating a visible and accessible archive of lived interaction with health technologies. A selection of these contributions could then be included in the upcoming IH conference programme and, where appropriate, incorporated into formal proceedings through ACM publication channels, with authors invited to present or participate in sessions. This pathway would, in turn, allow experiential knowledge to circulate both in conversational spaces and within formal archival structures, linking visibility, dialogue, and the scholarly record.

\section{Other Possible Views}

From a positivist standpoint, experiential accounts may be understood as data to be analysed and generalised in order to produce reliable, comparable, and cumulative knowledge. Within this view, transforming lived experience into themes or models is appropriate and necessary for establishing shared evidence and advancing the field. This position has value in enabling comparability, scalability, and methodological rigour. Meanwhile, such transformations may also abstract away the temporal, relational, and affective dimensions through which interactive health systems are lived. In doing so, experiential knowledge risks being reduced to what can be stabilised and formalised within existing analytic frameworks, rather than being recognised in forms that reflect its situated and ongoing nature. 

From this perspective, similar issues arise in relation to authorship, which entails a different form of epistemic and institutional recognition. However, this does not straightforwardly translate into a desirable or appropriate role for all participants. The dynamics through which experiential accounts become research data, and the power relations embedded in researcher-participant interactions, are more complex and situated than a simple shift from acknowledgement to authorship might suggest. Therefore, the views and proposal outlined in this paper do not seek to replace these research practices, but to complement them by introducing pathways through which experiential knowledge can be expressed and shared as knowledge in its own right.

\section{Future Directions}

I recognise that this proposal does not yet address several practical and structural questions. For instance, creating a formal space does not, in itself, ensure that diverse patient communities are aware of it, able to access it, or willing to contribute to it. In many existing venues mentioned previously, first-person sections tend to be populated by contributors who are already comfortable navigating academic publication systems, including patients who are themselves researchers or health professionals. Therefore, I end this paper by asking the question: how might such opportunities extend beyond those already proximate to academia? 

If IH were to explore such a structural pathway, one of the next steps would be to prototype it in practice, while explicitly attending to outreach, accessibility, and sustained engagement beyond established academic networks. For example, we could pilot an experiential call alongside an upcoming IH conference cycle, experiment with lightweight submission routes, or curate a small set of contributions for blog, zine, or programme inclusion. Observing how these contributions are reviewed and taken up would generate practical insight into what forms of guidance, review, and stewardship are needed. As an initial step in this direction, I have created an accompanying online space where readers can engage with and respond to the ideas presented in this paper, as well as contributions from others: \url{https://yuhaosun.com/patient-voice/}. Additionally, bringing together interested members of the IH community through formats such as workshops, meet-ups, or Birds of a Feather\footnote{The issues of patient voice and lived experience discussed in this paper are also being explored in a Birds of a Feather session at IH 2026 on ``Precision Medicine in Practice: What Role for HCI?'' \cite{sun2026role}, where related questions are taken up in a lightweight and exploratory format as part of ongoing efforts to engage the IH community.} sessions may provide an initial space for collective exploration.

More broadly, I invite the field of IH to reflect on how its knowledge infrastructures are designed. The question is not only whether patients are engaged in research processes, but how experiential accounts are positioned within the scholarly record. Treating this as a design problem opens space for experimentation, comparison, and revision. As the IH community continues to study and design interactive health systems, expanding how it learns from those who live with them may prove as consequential as the systems it builds. Ultimately, the question is how the field of IH designs the epistemic structures through which those voices can shape what counts as knowledge.

\begin{acks}
This work was completed during my recovery from cholesteatoma surgery. Throughout this time, I found strength in many `patient voice' articles written by fellow patients. I am grateful for their openness in sharing such powerful and encouraging stories. I thank the anonymous reviewers for their helpful comments. 

My thanks also go to my mother, my father, and my fiancé for their love during difficult moments. I owe them more than words can express.

\end{acks}

\bibliographystyle{ACM-Reference-Format}
\bibliography{sample-base}


\end{document}